\documentclass{aa}  
\usepackage{graphicx}
\usepackage{txfonts}

\begin{document}

   \title{Convective cells in Betelgeuse: imaging through spectropolarimetry\thanks{Based on observations obtained at the T\'elescope Bernard Lyot
(TBL) at Observatoire du Pic du Midi, CNRS/INSU and Universit\'e de
Toulouse, France.}}

   \author{{\sc A.~L{\'o}pez Ariste}\inst{1}, {\sc P. Mathias}\inst{1}, {\sc B. Tessore}\inst{2},  {\sc A. L\`ebre}\inst{2}, {\sc M. Auri\`ere}\inst{1}, {\sc P. Petit}\inst{1},  {\sc N. Ikhenache}\inst{1},  {\sc E. Josselin}\inst{1,2}, {\sc J. Morin}\inst{2}, {\sc M. Montarg\`es}\inst{3}.}
   \institute{IRAP, Universit\'e de Toulouse, CNRS, CNES, UPS.  14, Av. E. Belin. 31400 Toulouse, France \and
LUPM, Universit\'e de Montpellier, CNRS, Place Eug\`ene Bataillon, 34095 Montpellier, France \and
Institute of Astronomy, KU Leuven, Celestijnenlaan 200D B2401, 3001 Leuven, Belgium}
   \date{Received ...; accepted ...}

 
  \abstract
  {}
{We assess the ability to image  the photosphere of red supergiants and, in particular Betelgeuse, through the
modelling of the observed linear polarization in atomic spectral lines.
We also aim to  analyse  the resulting images over time, to measure the size and dynamics of the convective structures in
these stars.}
{Rayleigh scattering polarizes the continuum and spectral lines depolarize it.
This depolarization is seen as a linear polarization signal  parallel to the radial direction on the stellar disk.
Integrated over the disk, it would result in a null signal, except if brightness asymmetries/inhomogeneities are present.
This is the basic concept behind our imaging technique.
Through several tests and comparisons, we have tried to assess and extend its validity, and to determine what can be learnt
unambiguously through it.}
{The several tests and comparisons performed prove that our technique reliably retrieves the salient brightness structures
in the photosphere of Betelgeuse, and should be relevant to other red supergiants.
For Betelgeuse, we demonstrate that these structures we infer are convective cells, with a characteristic size of more than 60\,\% of
the stellar radius.
We also derive the characteristic upflow and downflow speeds, 22 and 10\,km.s$^{-1}$ respectively.
We find weak magnetic fields concentrated in the downflow lanes in between granules, similarly to the quiet sun magnetism.
We follow those convective structures in time.
Changes happen on timescales of one week, but individual structures can be tracked over  four years of  observations.}
{The measured characteristics of the convection in Betelgeuse confirm the predictions of numerical simulations in both the strong,
supersonic upflows as well as the size of the convective cells.
They also concur in the presence of  weak magnetic fields that are completely dominated by the convective flows and constrained to the
dark lanes of down-flowing plasma.}
  

\titlerunning{Convective cells in Betelgeuse}
\authorrunning{A. L\'opez Ariste et al.}

   \maketitle

\section{Introduction}

The chemical enrichment of the interstellar medium is mostly due to the late stages of stellar evolution;
this is particularly true for red supergiants (RSG) because of their mass loss.
However, the mechanisms that drive mass loss from these stars are not well understood.
Mechanisms that are often invoked include thermal gas and radiation pressure, acoustic and shock waves, Alfv\'en waves, magnetism,
and most probably other additional phenomena.
The strong wind of these stars is probably triggered by dust, formed at a great distance from the photosphere, and so has
 no direct connection with the atmosphere dynamics.
It is thus essential to characterize the phenomena that take place close to the stellar surface.
\cite{Josselin2007} suggested that high velocities and steep velocity gradients, possibly caused by convective motion,
generate line asymmetries, that turbulent pressure decreases the effective gravity, and that this decrease combined with
radiative pressure on lines initiates the mass loss.
Convection seems indeed to be a key component to understand the evolution of massive, cool evolved stars.
\cite{Schwarzschild1975} suggested that the outer envelope of RSG could host a small number of large convective cells.
These gigantic convection cells have also been predicted in simulations \cite[e.g.,][]{Freytag2002,Chiavassa2011}
and suggested in observations of Betelgeuse as bright photospheric spots observed with direct UV imaging \cite{Gilliland1996} or with interferometric techniques \cite[e.g.,][]{Haubois2009}.

Indeed, because of its relative proximity (about 200\,pc), and as an MIab supergiant, Betelgeuse offers the largest
angular diameter of any  star (except for the Sun), and has been extensively studied in interferometry
\cite[e.g.,][]{Haubois2009,Montarges2016}.
Because interferometry may suffer from  modelling hypothesis  (limb-darkening, spot(s) shape...) and cannot resolve velocity fields, it is interesting to
develop alternative approaches that complement and reinforce these results.

Since 2013 the instrument Narval@TBL has been monitoring linear polarization on Betelgeuse with roughly one observation per month 
during the visibility period of the year, leading up to 43 observations in total through April 2018. 
\cite{Auriere2016} first interpreted linear polarization in terms of Rayleigh scattering in the continuum and depolarization 
during the formation of spectral lines. 
Those authors also proposed a first modelling of linear polarization through bright spots that would result in the net linear polarization we observe. 
A tentative confirmation of such bright spots, in addition to the linear polarization interpretation, was seen in the images inferred from 
interferometry with the PIONEER instrument \cite{Montarges2016}. 
Beyond that first success, it was already obvious that the profiles could not be interpreted as simply being due to one or two bright spots on top of 
an otherwise homogeneous photosphere. 
Further insight came from \cite{Mathias2018} who added to the linear polarization the analysis of the circular polarization signatures, due to the Zeeman effect, 
monitored over more than eight years.
Two results were crucial for two reasons. 
First, the signals presented a rough periodicity, as seen in the Fourier power maps, of about 1800 days, a timescale usually associated with the convective 
one for a red supergiant as Betelgeuse \cite{Stothers2010}. 
Second, the linear polarization signals were predominantly found blue-shifted with respect to the intensity signals, while Zeeman-induced circular polarization was red-shifted. 
Both results pointed towards a conclusion that is the subject of the present paper: the bright spots identified through linear polarization were the convective 
cells of Betelgeuse seen at the photospheric level whereas the tiny magnetic fields were dragged around by the convective flows and were thus concentrated in the 
intergranular dark lanes where the plasma sinks into the stellar interior. 

The path towards imaging the photosphere of Betelgeuse is based upon the interpretation that we are seeing convective cells. 
This allows us to propose models of the brightness of the photosphere of Betelgeuse in terms of spherical harmonics and to link the 
Doppler shift of the local spectral profile to this brightness. 
This is introduced in quite some detail in Sect. 2. 
The model parameters are the coefficients of the decomposition of the brightness in terms of spherical harmonics. 
We present in Sect. 3 a simple inversion algorithm based upon the Marquardt-Levemberg iteration that allows us to fit the observed linear polarization 
profiles in terms of an image of the photosphere of Betelgeuse. 
In Sect.\,4, we present the data that allowed us to conclude on the convective model and which provide some preliminary estimates of the velocities involved 
in the convection of Betelgeuse. 
The size of the convection cells is analysed in Sect.\,5, and a characteristic length scale of $0.62\,R_*$ is found for these structures. 
Finally, in Sect.\,6, we discuss the four years of convective dynamics of Betelgeuse. 
Most of the data presented here have already been described by \cite{Auriere2016} and \cite{Mathias2018} but, since, the monitoring observations of 
Betelgeuse continues, with new observations corresponding to 2017-2018. 

\section{Origin and modelisation of linear polarization profiles}
\subsection{Observational facts and previous basic model}

At high polarization sensitivity, the atomic photospheric lines in the spectra of Betelgeuse show linear polarization signals. 
\cite{Auriere2016} were the first to show these signals over three years of observations of Betelgeuse.
In this paper we present the continuation of this series of measurements up to April 2018. Since then linear polarization signatures have been detected in the spectra of other red supergiants \cite{Tessore2017}.

\begin{figure}[htbp]
\includegraphics[width=0.5\textwidth]{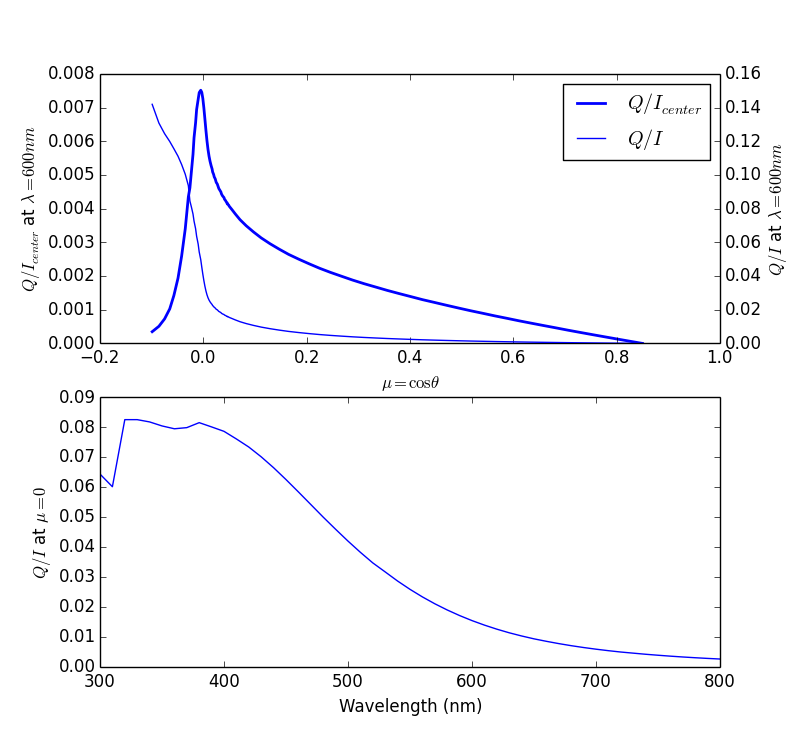}
\caption{Expected linear polarization rates in the continuum due to Rayleigh scattering in a model atmosphere adequate for Betelgeuse
($T_{eff}=3750K$, $\log g=0.5$, $M=15 M_{sun}$, $v_t=5 km s^{-1}$ and solar composition). Top: Polarization rates at the fixed wavelentgh of 600nm as a function of the angular distance $\mu =\cos \theta$ to disk centre. The envelope, beyond the stellar radius, is artificially shown with negative values of $\mu = \cos \theta$. The thin line gives the actual polarization rate respect to the local intensity. The thick line gives the rate normalized to the intensity at the disk centre, providing a  measure of the relative contribution to polarization the integrated polarization signal. Bottom: Dependence of the polarization rate with wavelength at a fixed distance to disk centre $\mu =\cos \theta=0.$.}
\label{J20}
\end{figure}

The amplitude of those signals is smaller than 0.1\,\% of the continuum intensity and, characteristically, several peaks with different signs 
are seen across and sometimes beyond the width of the intensity profile.  
\cite{Auriere2016} produced the basic explanation of the origin of that signal that we now summarize. 
The photospheric continuum in Betelgeuse is polarized by Rayleigh scattering. 
The local amplitude of this polarization can be estimated using a MARCS atmospheric model for Betelgeuse 
with  $T_{eff}=3750K$, $\log g=0.5$, $M=15 M_{sun}$, $v_t=5km s^{-1}$ and solar composition \cite{Gustafsson2008}. The wavelength and position dependence of this Rayleigh polarization is illustrated in Fig.\ref{J20}.  With respect to the integrated intensity, the maximum contribution comes from regions at  $\mu=\cos \theta=0$ (with $\mu=1$, $\theta =0$ corresponding to disk centre and $\mu =0$, $\theta=\pi/2$ to the photospheric limb, where the line of sight is perpendicular to the stellar radius) and, at this distance from disk centre, increases from about 1\% in the red to up to 8\% in the blue. These numbers update and correct previous estimates \cite[e.g.][] {Doherty1986}, mostly because of the improved atmospheric model. The polarization amplitudes so computed are  about 100 times larger than the corresponding ones 
measured in the limb of the Sun \cite{Leroy1972,Stenflo2005,Fluri1999} at similar position and wavelength. 
This polarized continuum is then absorbed by atoms in the regions above the continuum forming region. 
The de-excitation of these atoms leads to a decrease of the local polarization since the photon, once absorbed, is simply re-emitted unpolarized. 
Thus, atomic spectral lines depolarize the continuum Rayleigh polarization when forming in the atmosphere of Betelgeuse.

On top of this mechanism that is common to all spectral lines, specific lines may be sensitive to atomic alignment induced by the anisotropy of the radiation field. 
For those specific lines, the re-emitted photons are polarized. 
The details of these two competing mechanisms can be found in \cite{LandiDeglInnocenti2004}. 
However,  \cite{Auriere2016}  demonstrated that the spectra of Betelgeuse were compatible with a scenario in which, unlike the solar case,  the intrinsic line 
polarization was negligible, all lines just depolarizing the continuum.  The observation of the linear polarization in the \ion{Na}{i} $D_1$ and $D_2$ lines was the key to this conclusion. While both lines are equally able to depolarize the continuum (an ability that mostly depends on the depth of the line), the $D_2$ line produces 10 to 100 times more intrinsic polarization than the $D_1$ line. Unlike the solar spectrum, where the $D_2$ line is 10 times more polarized than $D_1$, in Betelgeuse both lines showed a very similar signal both in amplitude and shape, an observation only compatible with the depolarization process. This conclusion could be easily extended to most other lines, with smaller intrinsic polarization levels thus leading to the conclusion that all lines were depolarizing the continuum, unlike the solar case.
The  reasons behind the differences between this scenario in Betelgeuse and the solar case are left for future work.
In these conditions, were we  able to observe a local point at the limb of Betelgeuse, we would see a polarized continuum along the direction 
tangent to the local limb. Spectral lines that depolarize this continuum would decrease this polarization signal at their respective wavelengths. 
This local polarization spectrum would look very much like the intensity spectra.

However, in an unresolved star, the observed global polarized spectrum is integrated over the whole stellar disk. 
Since the linear polarization direction rotates over $2\pi$, linear polarization signals should cancel out, except if brightness or polarization 
inhomogeneities are present over the stellar disk. 
This is the basic idea behind the techniques of imaging through spectropolarimetry.  
The ratio of Stokes $Q$ to Stokes $U$  in the continuum could eventually provide the polar angle at which the maximum brightness is found. But this is hardly sufficient for imaging and, furthermore, our instruments do not measure the polarization in the continuum   
(while relative polarization, e.g. the polarization in the line respect to the continuum, is measured down to sensitivities of 1 in $10^{5}$, absolute polarization, as measuring the continuum polarization requires, is limited by photometric and hardly goes beyond sensitivities of 1 in  1000).
The situation is improved by the analysis of the polarization signals within the spectral lines. 
The observed signals inside atomic lines are found at different Doppler shifts. 
Assuming a fixed radial velocity, the Doppler shift of each signal provides a distance to the disk centre. 
Simultaneously, and as for the continuum, the ratio of $Q$ to $U$ Stokes parameters fixes the polar angle, but this time we obtain one angle per signal.
Thus, the combination of Doppler shift, due to a hypothetically constant radial velocity, with the ratio $Q/U$, due to the depolarization of the continuum 
with polarization tangent to the local limb, leads to a two dimensional mapping of brightness inhomogeneities over the stellar disk.

As the polarization of all lines is due to continuum depolarization, the spectral shape of this polarization is that of the spectral line and proportional to the depth of the line. Under these circumstances, we can add up several thousand lines over the spectra after scaling wavelength to a velocity respect to line centre for each line and weighting the signals by the depth of the intensity line. This is a basic description of LSD  \cite{Donati1997} which will effectively increase the signal-to-noise ratio of the resulting pseudo-line.
Applying the previous basic mapping schemeto the LSD pseudo-line, \cite{Auriere2016} computed for each wavelength position across the spectroscopic line the polar angle at which brightness was maximum 
and plotted that as a spot. 
The resulting maps consistently show two to three bright spots. 
At two particular dates, Jan 9th, 2014 and Dec 9th, 2015, interferometric observations were available that concurred on the presence of 
hot spots at at least one of the  selected positions and up to a 180 degrees ambiguity in the azimuth angle of the spots, 
inherent to the polarization measurements  \cite{Montarges2016,Auriere2016}.
Despite  its crudeness, this basic method was able to pinpoint some of the brightest spots over Betelgeuse.

Among the many simplifications onto which a simple first such model is based, one drives our present improvement in modelling the linear polarization of Betelgeuse. 
At any distance of the disk centre, determined by the Doppler shift of the polarization signal, the basic method assigns the polar angle to a unique source 
of brightness, independently of other sources. 
However,  it is obvious that the combination of several spots may result in a  signal peak at the correct wavelength and with the right $Q/U$ ratio.  
But how could one find the information to localize two spots that combine their signals into 
a single maximum of linear polarization? 
While at first sight it appears that such improvement is beyond hope, further thought reveals that such information exists in the data: if the bright spot was a single, isolated one, both $Q$ and $U$ would produce a maximum at the same wavelength. But
this is not always the case, as can be seen by inspecting in detail the profiles shown in \cite{Auriere2016}, \cite{Mathias2018}. 
The maxima of $Q$ and $U$ often occur at different wavelengths. 
This indicates the presence of  ad minima asymmetrical spots and, in general, of multiple spots. 

 \cite{Mathias2018} adds a second, indirect argument to that described above. 
In that work, \cite{Mathias2018} present simultaneous circular and linear measurements. 
The Stokes $V$ circular polarization is shown to be due to the Zeeman effect and attributed to photospheric magnetic fields with net longitudinal 
fluxes of 1\,G or less. 
In addition, most of the Zeeman signatures are redshifted, while the linear polarization signatures are predominantly blueshifted (see also Sect.\,5). 
The basic, simplest interpretation of these observations is the same that explains the similar behaviour of the very weakest fields found in the solar photosphere.
These weak fields are unable to impact the plasma dynamics and are therefore dragged by the main convection flows over the surface. 
Hot plasma rises in the middle of hot granules and moves horizontally 
 to the granular edges while cooling; 
during this advective motion, it drags any weak magnetic field into the intergranular dark lanes where the plasma sinks. 
Plasma flows converge into these intergranular lanes and magnetic fields dragged into those places cannot escape. 
In summary, the brightest places that, we assume, would be the largest contributors to the net linear polarization correspond to the centre of convection 
cells and are blue-shifted, whereas magnetic fields are  expected at the intergranular dark lanes where plasma sinks into the star and are thus redshifted. 
This corresponds neatly to the observational data from the polarization signals in Betelgeuse. 
We must conclude that linear polarization in Betelgeuse comes from the rising plasma in the centre of convection cells and that these convection cells cover the surface of Betelgeuse, leaving only narrow, cold, redshifted intergranular lanes in which weak magnetic fields are found.

In view of this evidence, the surface of Betelgeuse cannot be modelled with a few bright spots on top of a homogeneously bright surface, but rather as a 
collection of convective cells with bright broad centres and dark narrow lanes.

\subsection{The new model}

Mathematically, the best way to do this modelling is by describing the surface brightness of Betelgeuse as the absolute value of a linear combination of real 
spherical harmonics up to a maximum value of the index $\ell$. 
By real spherical harmonics we mean that we handle   the real and imaginary parts of the usual complex spherical harmonics separately and with different coefficients.

\begin{figure*}[htbp]
\includegraphics[width=\textwidth]{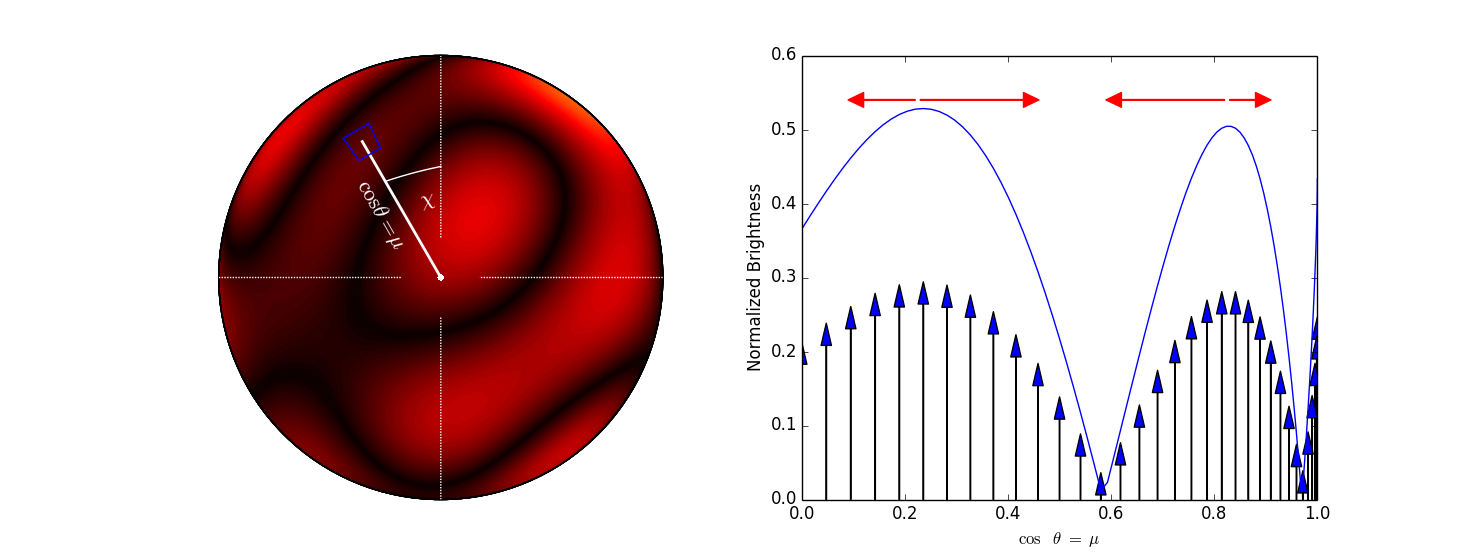}
\caption{ Random combination of spherical harmonics up to degree 5 represent the brilliance over the stellar disk. Left: Graphical definition of the coordinates of the blue square over the disk. Right: Brilliance of the disk along that particular radius in the left image. The vertical arrows indicate the vertical, radial, velocity included in the model to represent the emerging plasma velocity. We note that the zero velocity is shifted so that all velocities are positive. The red horizontal arrows represent the velocities of advection of the plasma towards the cold, integranular regions where it sinks down into the stellar interior. These horizontal velocities are NOT included in the actual model.}
\label{cartoon}
\end{figure*}

Specifically, the star brightness model is defined by a set of parameters $a_n$ which  result in  the brightness at a point at distance $\mu$ from disk centre
and  polar angle $\chi$ (see Fig.\ref{cartoon}) given by:
\begin{equation}
B(\mu,\chi)=\left \| \sum_{\substack{\ell=0,\ell_{max} \\ m=-\ell,+\ell}} a_{\ell}^m y_{\ell}^m(\mu,\chi) \right \|
\label{Beq}
\end{equation}

For completeness, we have defined the function $y_{\ell}^m(\mu,\chi)$ as
\begin{eqnarray}
y_{\ell}^m(\mu,\chi)=\sqrt{2} (-1)^m \Re(Y_{\ell}^m(\mu,\chi))  \quad \textrm{if} \quad m>0 \\
y_{\ell}^m(\mu,\chi)=\sqrt{2} (-1)^m \Im(Y_{\ell}^m(\mu,\chi))  \quad \textrm{if} \quad m<0 \\
y_{\ell}^m(\mu,\chi)=\Re(Y_{\ell}^m(\mu,\chi)) \quad \textrm{if} \quad m=0 
\end{eqnarray}
with $\Re$ and $\Im$ meaning the real and imaginary parts respectively, and, as usual,
\begin{equation}
Y^m_n(\mu,\chi) = \sqrt{\frac{2n+1}{4\pi} \frac{(n-m)!}{(n+m)!}}  e^{i m \chi} P^m_n(\mu)
  \end{equation}
where $P^m_n(\mu)$ is the associated Legendre polynomial.

The velocity at each point over the disk is radial and proportional to the brightness.  In the presence of convection, this proportionality can be justified by arguing that the radiative losses, hence the brightness, have to be roughly balanced by the enthalpy flux (see, e.g., \cite{Nordlund2009} discussing the solar granulation case). Brightness therefore acquires a dependence also on $v_z$ as
\begin{equation}
B(\mu,\chi,v_z) \propto \sigma T_{eff}^4 \approx \rho v_z \left( \frac{5}{2}k_BT+x\Xi \right)
\end{equation}
where the left side recalls the black body relationship between brightness and effective temperature, and on the right side, 
the gas temperature $T$ and the ionization potential of hydrogen $\Xi$ times the ionization fraction $x$, 
times the density $\rho$ are transported upwards with a vertical velocity $v_z$.
Our brightness distribution is adimensional and normalized. 
To provide a velocity with the right units but proportional to this brightness we define a global velocity $v_0$ that gives the maximum upflow velocity 
at the brightest point. 
This velocity $v_0$ must be large enough in order for the maximum Doppler shift (at disk centre) to encompass the most blue-shifted polarization signal  
observed in the  series of data. 
For Betelgeuse, this velocity is at least 40\,km.s$^{-1}$, but is rather 60\,km.s$^{-1}$ for $\mu$\,Cep and just 30\,km.s$^{-1}$ for CE\,Tau,  two comparative M-type RSGs for which we also have spectropolarimetric Narval observations \cite[linear and circular polarization, see ] []{Tessore2017a}. 
All put together, the Doppler velocity at any particular point over the disk will be modelled as 
\begin{equation}
v=v_0\cos(\mu) B(\mu,\chi,v_z)
\end{equation}
The dark intergranular lanes are assigned here a near zero velocity, rather than a negative one. 
This only offsets the zero velocity definition. 
This assumption that the brightness distribution corresponds to convective cells (and hence that velocity and brightness are related) is demonstrated in Sect. \,4. 
Finally, even if present,  horizontal velocities are not considered in our model.

We also have to fix the Doppler width of the local profile from a single point in the atmosphere of Betelgeuse. 
This width is actually the convolution of the physical broadening of the profile emitted by Betelgeuse (combination of the thermal broadening, micro- and 
macroturbulence and any other physical broadening mechanism) with the PSF of our spectrograph. 
This combined width is found again by examination of the data. 
The individual peaks in the polarization profiles, as well as the Stokes $I$ profiles, are found to have an average FWHM of 20\,km.s$^{-1}$ in our data. 
This width is probably the combination of the instrumental width and  various sources of stellar broadening that will limit our ability 
to resolve stellar structures. 
In an  ad hoc attempt to separate these two sources of broadening, we set the FWHM of a point source in our model to 10\,km.s$^{-1}$. 
The profile emitted by each point is  therefore described by a Gaussian of FWHM 10\,km.s$^{-1}$ centered at $v_z$:
\begin{equation}
e^{-(v-v_z)^2/6^2}.
\end{equation}
Integrated over the stellar disk, the total polarization profile is just the addition of these profiles for all points with amplitude given by 
the brightness weighted by the Rayleigh scattering geometry factor $\sin^2 \mu$ and projected into Stokes $Q$ and $U$ for the position angle
\begin{eqnarray}
Q(v)=\sum_{\mu,\chi,v_z} B(\mu,\chi) \sin^2 \mu \cos 2\chi e^{-(v-v_z)^2/6^2}\\
U(v)=\sum_{\mu,\chi,v_z} B(\mu,\chi) \sin^2 \mu \sin 2\chi e^{-(v-v_z)^2/6^2}
\label{QU}
\end{eqnarray}

\subsection{Last scattering approximation}
The simple addition of profiles is justified by assuming that the polarization is fully due to the last scattering of the photon on each point over the disk.
This assumption has been used in the past to model the continuum polarization of the Sun (Stenflo 2005) or to attempt to interpret the details of the 
second solar spectrum \cite[e.g., ][]{Anusha2010,Belluzzi2011}. Applied to the Sun or to the present case of Betelgeuse, this approximation assumes that
the degree of linear polarization is weak everywhere so that instead of being created through multiple scattering events, the emergent linear polarization 
at each point of the stellar surface is created by only one scattering event, the last one. 
Through this assumption, the only contribution of limb darkening onto the polarization signal is through the amplification  of the anisotropy in the radiation field at the height of scattering.

We also  hypothesise that the amplitude of linear polarization is determined uniquely by the brightness. 
This is of course not rigorously true. 
Locally, the amount of polarization in the continuum will be given by the local brightness, but also by the anisotropy of the radiation field on 
which, for example, the height of the scattering point over the atmosphere of Betelgeuse plays a non-negligible role. 
Thus, under the symbol $B(\mu,\chi,v_z)$, referred to as brightness above, we may also find height variations in the atmosphere. 
Only direct comparisons of our inferred images with those obtained by other techniques permit the assumption that actual brightness dominates  our adimensional parameter $B(\mu,\chi,v_z)$.


\section{Imaging through inversion}
\begin{figure*}[htbp]
\includegraphics[width=\textwidth]{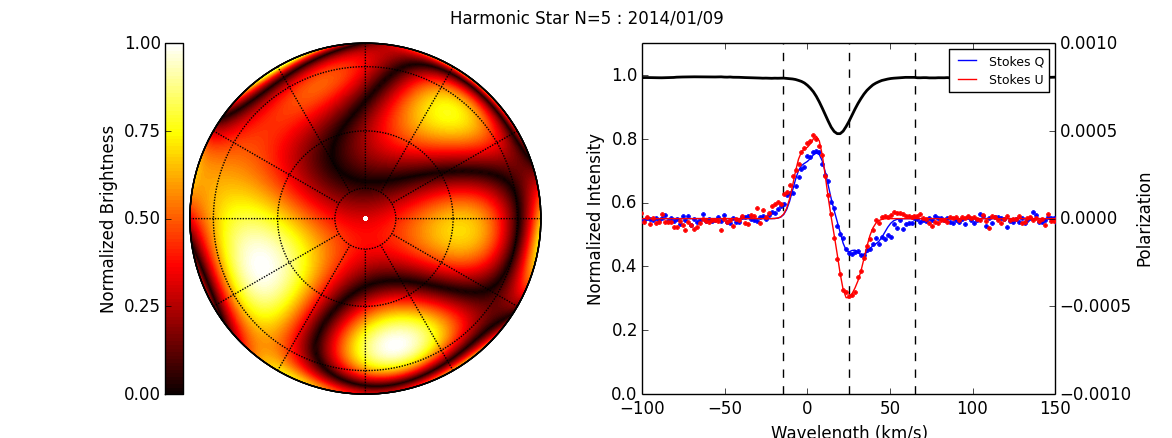}
\caption{Example of inversion. The LSD profiles for Stokes $Q$ and $U$, observed on January 9th, 2014, are shown as blue and red dots respectively at right. 
The observed intensity profile is also shown in black with ordinates at left. 
The inversion gave as solution the star shown at left which produces the profiles shown as continuous line at right.  The dotted lines over the image are plotted to help identify the polar angle and distance to centre coordinate system over the visible hemisphere of the star.
Over the spectra on the right, the vertical dotted lines indicate the centre of mass  velocity of the star, as well as  the blue and redshifted extreme velocities considered by the model. }
\label{plotSp_0109}
\end{figure*}

Given a set of Stokes $Q$ and $U$ profiles as those published by \cite{Auriere2016} or \cite{Mathias2018}, our purpose is to find the right 
set of $a_n$ parameters in Eq.\ref{Beq} that better fit the observed profiles. 
A simple Marquardt-Levemberg minimisation technique suffices to find that optimal fit with no further regularisation than the squared difference between 
model and observation $\chi^2$.
Figure\,\ref{plotSp_0109} shows one example of observations plotted as dots for both $Q$ and $U$ and the best fit as a continuous line. 
The brightness image that produces the best fit profiles is represented on the left.

The quality of the fit allows one to conclude that the brightness distribution is certainly one of the possible brightness distributions on Betelgeuse, 
though perhaps not the only one. 
One obvious ambiguity is the 180-degrees symmetry inherent to linear polarization dependence on geometry visible in the double angles of Eq.\ref{QU}. 
The image in Fig.\,\ref{plotSp_0109} can thus be rotated by 180 degrees with totally identical resulting profiles.
It cannot be excluded that such 180-degrees ambiguities may also be at work in other more subtle manner. For example it can be seen  that any linear combination of the original image and its 180-degrees ambiguous counterpart may also be a solution.
Another source of multiple solutions is the choice of $l_{max}$, which in Eq.\ref{Beq} sets the maximum quantum number $l$ of the the spherical harmonics used in the decomposition of the brilliance.  This value sets the size of the smallest structures that the code will retrieve. Therefore we discuss it at length in Sect. 5. But a first bound can be set by noticing that, given the number of wavelength bins with polarization signal, the maximum number of degrees of freedom in the data must be around 30. This corresponds roughly to the number of coefficients for $l_{max}=5$ (35), and this sets a limit to the highest $l_{max}$ we can afford  with the present data.
One is left with the multiple solutions that $l_{max}$ values of one to five may provide.

\begin{figure*}[htbp]
\includegraphics[width=0.5\textwidth]{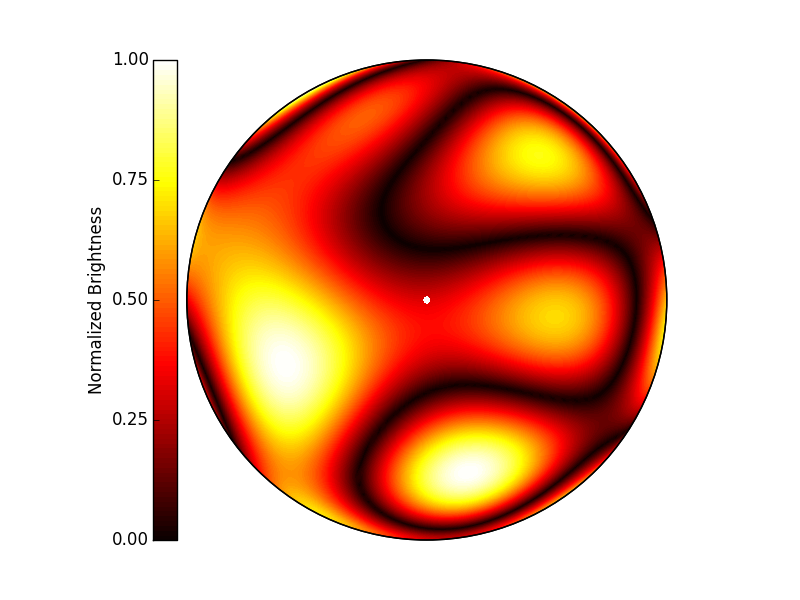}\includegraphics[width=0.4\textwidth]{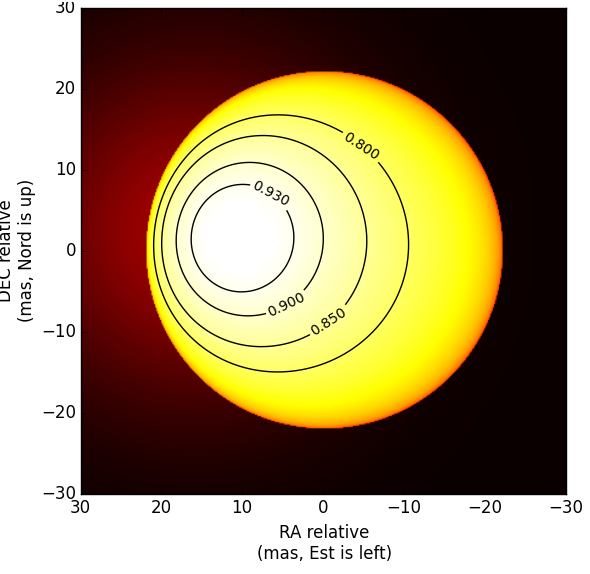}

\includegraphics[width=0.5\textwidth]{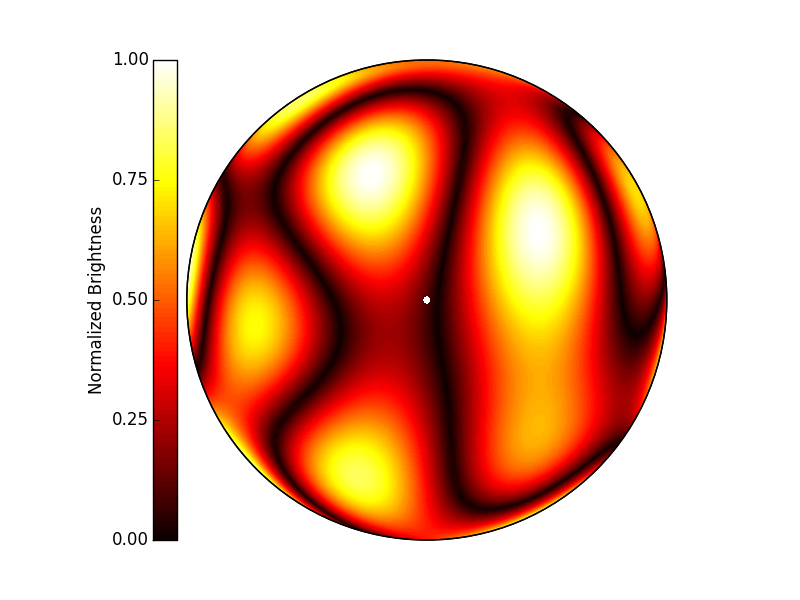}\includegraphics[width=0.4\textwidth]{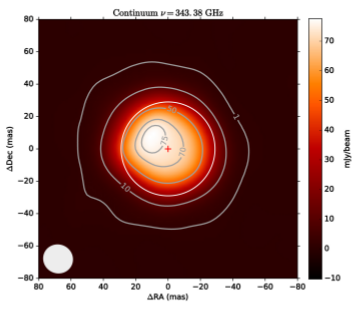}
\caption{Inferred images of Betelgeuse at January 9th,2014 (top left) and December 9th, 2015 (bottom left). 
The top left image corresponds in date with PIONEER observations published by \cite{Montarges2016} and \cite{Auriere2016} whose reconstructed image we reproduce at the top right. 
The bottom left image  can be compared with an ALMA observation made on November 9th, 2015 and published by \cite{Kervella2018} which is reproduced at the bottom right.}
\label{ALMA}
\end{figure*}

\begin{figure*}[htbp]
\includegraphics[width=0.5\textwidth]{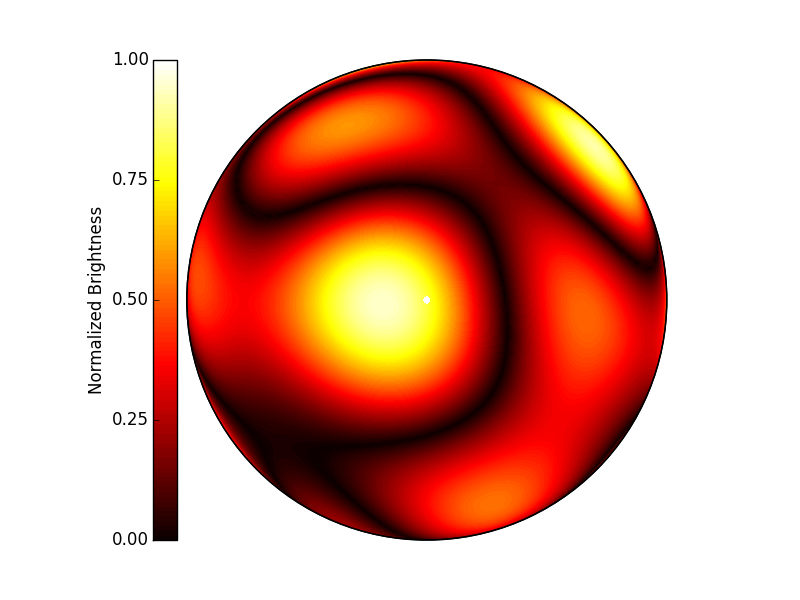}\includegraphics[width=0.5\textwidth]{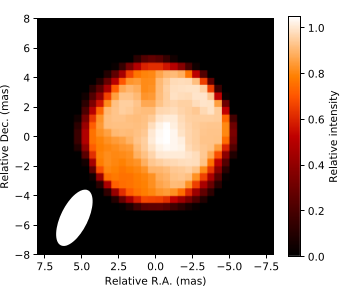}
\caption{Left: Image of CE Tau  on December 17th, 2016 as inferred from Narval spectropolarimetry. Right: Image of CE Tau on December 22nd-23rd, 2016 as reconstructed from CHARA interferometric data (adapted from Montarg\`es et al., 2018).}
\label{CETau}
\end{figure*}

Rather than a lengthy and cumbersome exploration of ambiguities and multiple solutions we have preferred to resort to comparisons of the images inferred with our present method with the
images obtained through other observational techniques as interferometry. 
Such comparisons are also produced by \cite{Auriere2016} and \cite{Montarges2016} between spectropolarimetric images with the basic model and 
 images inferred from the data obtained by the PIONEER interferometer at similar dates. 
Using the same data, we expect that the general agreement between our new model and their inferred images is still valid.  
Figure\,\ref{ALMA} (left) shows that this is the case for the date of January, 9th 2014 up to a 180 degrees rotation with respect to the solution shown in 
Fig.\,\ref{plotSp_0109}. 
More recently, on November 2015, ALMA also observed Betelgeuse with high resolution, though looking at high atmospheric layers \cite{Kervella2018}. 
Our data  from December 9th, 2015 shows  a correlation if one accepts that it is the brightest  north-east granule in our image (Fig.\ref{ALMA} right) that raises higher and thus 
becomes their single hot spot.
A final comparison has been made with  the other M type RSG, CE\,Tau, which has been observed by \cite{Montarges2018} with PIONEER on the VLT,  and which is also monitored in spectropolarimetry with Narval at TBL \cite{Tessore2017a} .
The reconstructed images from interferometry show a dominating central bright spot. 
Figure\,\ref{CETau} shows an image inferred from inversion of the linear polarization signals of this star with the same model described above except for the value 
of the $v_0$ velocity which  is reduced to 30\,km.s$^{-1}$ to better fit the observed Doppler span of the signals. 
A bright granule appears well centred in our image as well. 
This result is quite rewarding in two senses. 
First, a central granule is insensitive to the 180-degrees ambiguity.  Therefore, there are no ambiguous images that could also be solution of our inversion problem. The present image with the central granule is the only solution our model offers and it nicely corresponds with the results from interferometry.
Second,  since the amplitude of linear polarization due to scattering diminishes towards the disk centre, our technique is less sensitive to features near disk centre and will tend to disregard them. Despite  that, a dominant central granule is inferred.
Those two facts increase the confidence in the inference of such a central bright granule. 
The comparison with the interferometric observations is a final demonstration that our technique recovers the most salient features of the photospheric 
brightness distribution of these stars and, in particular, of Betelgeuse.

\begin{figure}[htbp]
\includegraphics[width=0.25\textwidth]{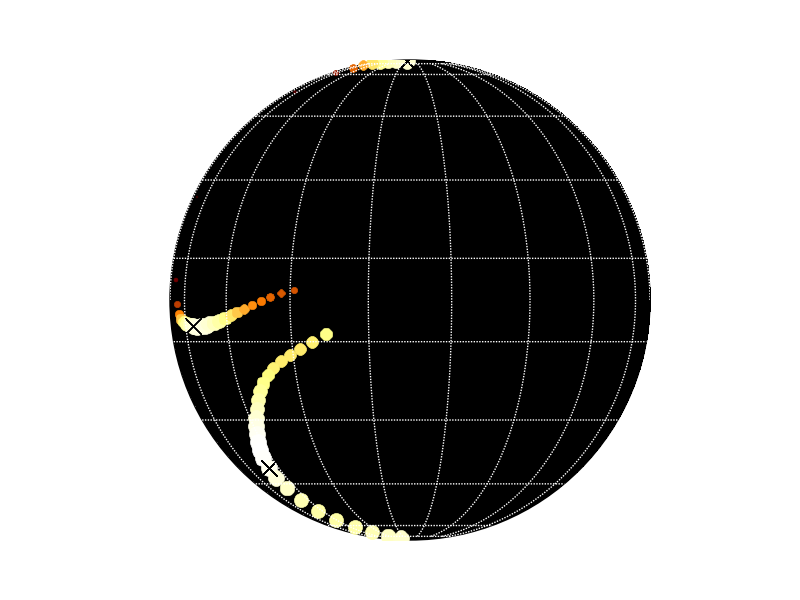}\includegraphics[width=0.25\textwidth]{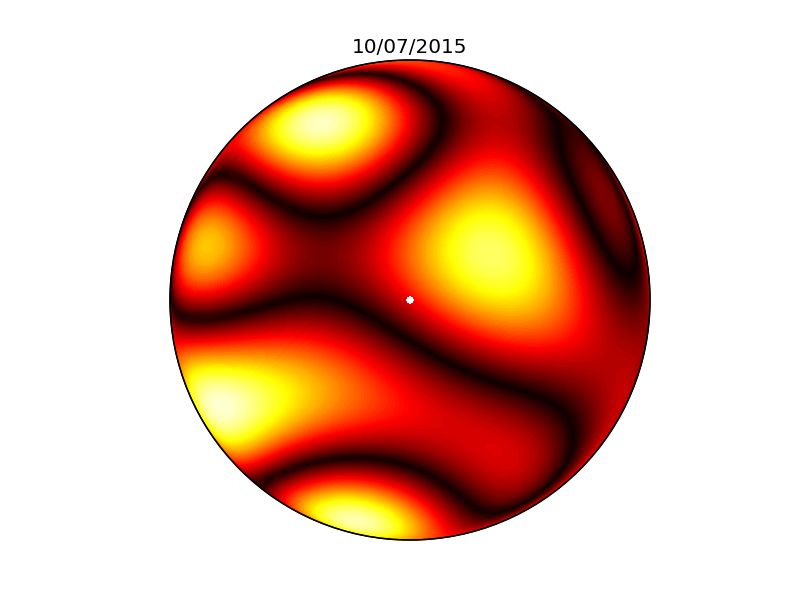}
\caption{Images of $\mu$\,Cep from July 10th, 2015. 
They are the result of the inversion of LSD profiles with the  basic model (left) or the new model based on spherical harmonics (right).}
\label{muCep}
\end{figure}

To determine if  modelling as complex as this in terms of spherical harmonics was necessary just to reproduce the main brightness features captured by other techniques, we refer the reader to Fig. \ref{muCep}. 
This single observation of $\mu$\,Cep on July 10th, 2017 is interpreted through the basic model of \cite{Auriere2016} and also with the new model based upon  spherical harmonics.
The original basic method suffers from the fact that Stokes $Q$ and $U$ profiles have peaks at quite different wavelengths. 
The only available interpretation of such misaligned peaks in the basic model is in terms of elongated structures. 
But they are difficult to explain or justify in stellar terms. 
However, the very same profiles interpreted with the present method result in a characteristic image of several bright spots of round shapes, 
in agreement with the interpretation of the brightness distribution as convective cells. 
This is one further confidence test supporting the conclusion that we are imaging true brightness structures in the photosphere of these supergiants.

The positive comparison of our inferred images with images retrieved through other techniques must be tainted, to end this section, with some negative remarks. 
The model presented above is able to reproduce all signals blue-shifted with respect to the  star reference. 
But no redshifted signal can be explained with our model. 
However, some observed profiles do show small red-shifted signals which, while showing amplitudes three times smaller than the blue-shifted ones, 
are shifted to the red with up to 15\% of the maximum Doppler shifts to the blue.
One possible explanation for these red-shifted signals is that they are coming from the dark intergranular lanes where plasma may be sinking back towards 
the stellar interior (we assume here the convective explanation of these bright spots that will be demonstrated in the next section). 
However these lanes are dark and occupy a small area over the star, so that their contribution to the integrated signal cannot be large. 
Another possibility is that those redshifted signals are due to horizontal velocities when seen near the limb. 
Horizontal flows such as advection must diverge from the centres of the granules with velocities of the order of the speed of sound ($\sim 6$\,km.s$^{-1}$).
Near the limb, those horizontal velocities would project onto the line of sight with positive or negative sign. 
The blueshifted projection would just add and modify the rest of the signal, and the redshifted projection would stand prominently with maximum 
redshifts comparable to the speed of sound. 
A third explanation is that the convective cells of Betelgeuse may rise high enough that we are able to see the top of those cells beyond the stellar limb. 
None of these explanations has been implemented in our model and thus those red-shifted signatures are left unexplained. 
Since they are small amplitude signals and completely independant from the rest of the signals, we argue that they are not affecting the images we present. 
However this may not be the case for all red supergiants. 
These signals are a clear indication that our simple model should evolve towards more sophisticated and realistic scenari.

\section{Convective cells}
The central result of this work is that we are imaging convective structures on the photosphere  of RSG through our spectropolarimetric technique. 
All the images shown in this paper's figures do indeed show what can be easily interpreted as large convective cells in both 
shapes and sizes expected for red supergiants. 
However, we can provide further evidence that this is the right interpretation of these structures. 

In convective cells, one expects hot bright plasma rising over the cell. As density drops, the hot plasma loses flotability and stops rising. Plasma movements are then due exclusively to any horizontal velocity present. As long as plasma is hot, and while it cools down through radiation, these horizontal movements will dominate the dynamics. Finally, the cold plasma  sinks in the dark, cold lanes between cells. 
These are the signatures that we must see in our data and images in order to confirm the convective nature of the imaged structures. 
To produce this evidence we must add to the picture the Stokes $V$ profiles observed often simultaneously with the Stokes $Q$ and $U$ profiles. 
These Stokes $V$ profiles were introduced and discussed by \cite{Mathias2018} who demonstrated also the Zeeman origin of these signatures. 
Through those signatures, the presence of weak magnetic fields in the photosphere of Betelgeuse was confirmed. 
A time-Fourier decomposition of these Stokes $V$ signals, observed since 2009, showed a quasi-periodicity of about 2000 days that has been linked 
to the characteristic convection scales \cite[e.g.][]{Stothers2010}.

In the convective atmosphere of the Sun one observes two regimes of magnetism. 
The strongest fields are able to maintain their topology against the convective motions of the plasma and even succeed in stopping convection altogether. 
They form active regions and sunspots. 
Weaker fields  cannot, however, impact plasma movements and are dragged around by the flows. 
They end up in the intergranular lanes and, in particular, in the network that marks the boundaries of supergranulation. 
These weak fields trace the borders between convective cells where the cold plasma sinks into the solar interior.  

\begin{figure}[htbp]
\includegraphics[width=0.5\textwidth]{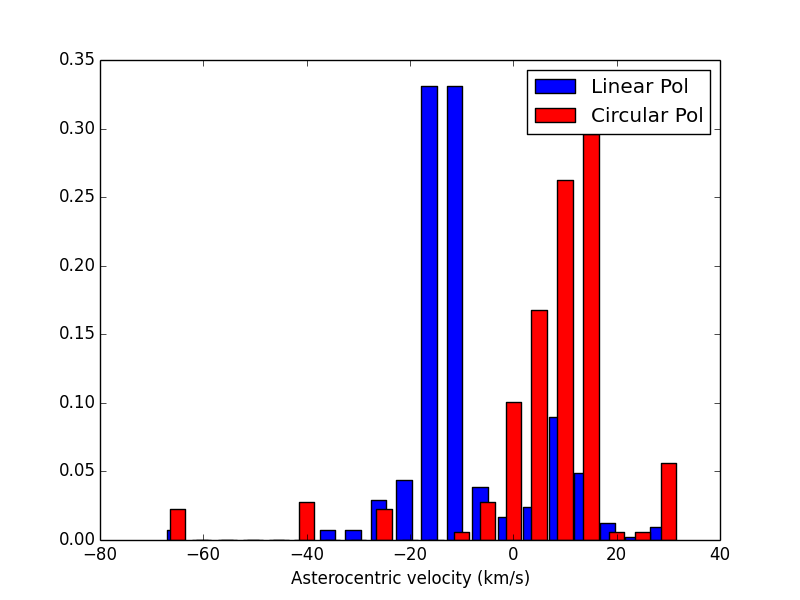}
\caption{Histograms of the Doppler velocities (respect to the centre of mass of the star) of the peaks in the profiles of linear polarization (blue) and circular polarization (red) for the whole 
dataset available for Betelgeuse (2013-2018). 
Each peak in the profiles has been given a weight proportional to its amplitude to stress the large signals over small, mostly noise-induced, peaks.}
\label{histogram}
\end{figure}
Figure\,\ref{histogram} shows histograms of the velocities of the linear polarization and of the Stokes $V$ peaks. 
Since the algorithm for the determination of peaks may be easily mislead by noise, we have weighted the histogram by the amplitude of the peaks
(a peak with an amplitude twice that of another peak is counted twice for the statistics). 
Linear polarization, arising from the brightest places, that is, the centre of the convection cells, is blueshifted by about 15\,km.s$^{-1}$ in about 70\,\% 
of the weighted cases. 
A secondary peak of linear polarization appears at +10\,km.s$^{-1}$, thus redshifted. 
This secondary peak corresponds to those low amplitude peaks left largely unexplained, as said in the previous section.
 
The Stokes $V$ signals on the other hand are redshifted by about 10\,km.s$^{-1}$. 
This may be easily compared with those weak fields in the solar case, that are dragged around by plasma motions and found in the 
 dark lanes between convective cells where the cold plasma sinks. 
Numerical simulations of magnetic fields in Betelgeuse \cite{Freytag2002} confirm this general scenario. 
We can interpret these histograms as follows.
The linear polarization comes from the brightest places, that is from the centre of convection cells where hot plasma rises into the atmosphere and 
blue-shifts the profile. 
Stokes $V$ comes from photospheric magnetic fields in Betelgeuse that would be too weak to alter plasma motions and  tend to concentrate in the 
intergranular lanes. 
Thus, Fig.\,\ref{histogram} confirms that the bright spots imaged with our technique in Betelgeuse are actual convection cells. 
And, by extension, those bright spots observed with other techniques and which coincide with our spots also are due to convective cells, though perhaps 
those that overshoot into higher atmospheric layers, depending on the observation wavelength.
 
The average velocities measured are around 15\,km.s$^{-1}$,  both blue and red-shifted. 
This must correspond to the typical radial velocities of convection in Betelgeuse weighted by the $\cos \mu$ integration factor over the stellar disk, 
or two thirds of the typical radial velocity of those convective movements.  
We have to conclude that the typical convective velocity in Betelgeuse is of the order of 22\,km.s$^{-1}$. 
This matches nicely with the velocities found by \cite{Freytag2002} in numerical simulation codes, velocities which are up to five or seven times the speed of sound. 
More classical values, as those of 7\,km.s$^{-1}$ found by \cite{Stothers2010} from basic convection theory, are too small to fit the profiles, and have 
to be understood as velocities occurring deeper in the convective interior, before the density and temperature drops convert the rise velocities into shocks.

\section{Size of convective cells}

\begin{figure}[htbp]
\includegraphics[width=0.5\textwidth]{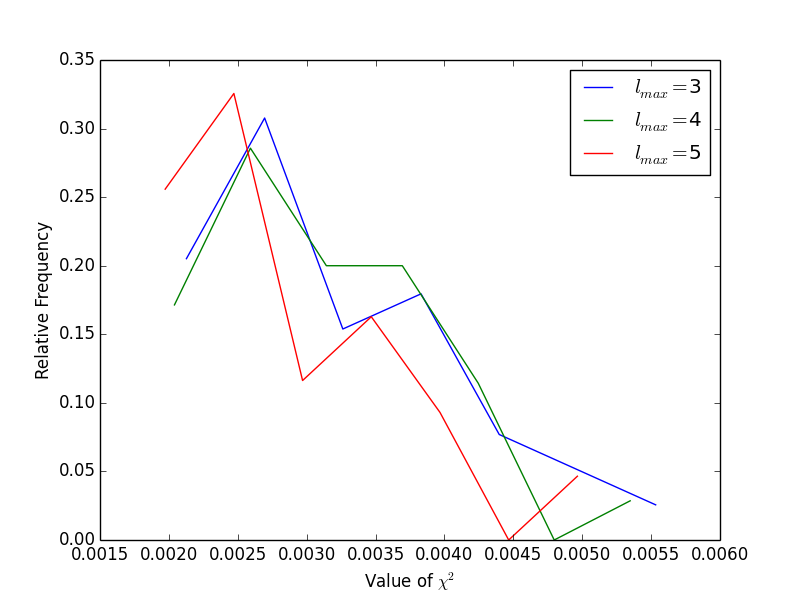}
\caption{Histograms of the final value of the merit function $\chi^2$ after inversion of the whole dataset of Betelgeuse with
$\ell_{max}=3,4,$ and 5 respectively. A marginal improvement of the inversion results is seen with larger values of $\ell_{max}$.}
\label{Chi2}
\end{figure}

The inversion code described in Sect.\,2.2 requires, among other parameters, to fix the maximum value of the quantum number $\ell$.
By fixing this number we constrain the size of the structures imaged on the surface of Betelgeuse or in other words, the size of the convective cells. 
Figure\,\ref{Chi2} shows a histogram of the merit function $\chi^2$ when inverting the full data set with $\ell_{max}=3,4,$ and 5 respectively.  
There is a marginal diminution of the merit function for $\ell=5$.

\begin{figure*}[htbp]
\includegraphics[width=1\textwidth]{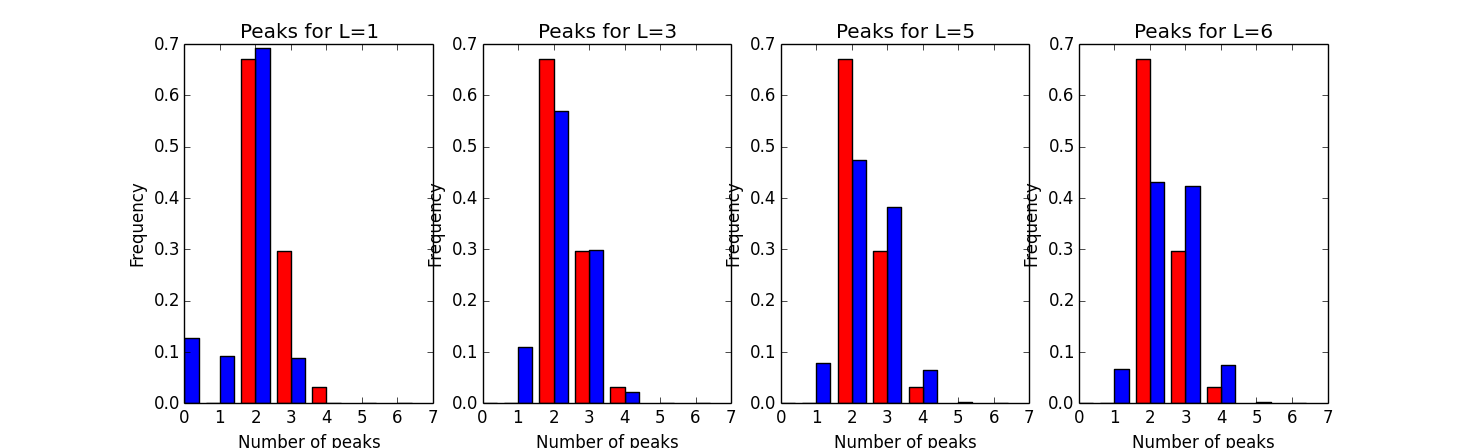}
\caption{In blue, histograms of the number of peaks in the profiles of linear polarization found from 1000 synthetic cases of stars with 
(from left to right) $\ell_{max}=1,3,5,$ and 6. 
In red, common to the four plots, the histograms of the number of peaks in the real dataset of Betelgeuse. 
The comparison of both histograms is satisfactory for $\ell_{max}=3$ to 5.}
\label{L16}
\end{figure*}

We found a better way to determine the right value for $\ell_{max}$ through the distribution of the number of peaks in the profiles as a function of $\ell_{max}$. 
The basic idea is that, in general, the larger $\ell_{max}$, the more granules will be visible in the image and, since each granule produces a peak in the 
profiles at the right wavelength, more peaks will be present in the profiles. 
This statement should however be tempered since misinterpretation may arise from the location of the granules on the stellar disk, the geometry of which
may possibly lead to mutual cancellation
if two granules merge into a single peak. Also, the spectrograph resolution may mask thin peaks
Keeping in mind all these arguments, we counted the peaks in both Stokes $U$ and Stokes $Q$ in the observed profiles. 
The histogram on the number of peaks is seen in red bars in the 4 plots of Fig.\,\ref{L16}.     

Next, we created random stars with $\ell_{max}\in [1;6]$ and computed the Stokes $Q$ and $U$ profiles arising 
from them in the framework of our new model. 
The histograms of the number of peaks in those sets of random stars are seen  in the blue bars of the  plots of Fig.\,\ref{L16}. 
Before comparing the observed and simulated histograms, we must keep in mind that the observations are a series of 43 dates over more than four years. Hence,  a temporal continuity is expected in that data set, and we shall not expect the observed histogram to represent the full dynamics of the convection of a star as Betelgeuse, but rather a snapshot. 
The counts from the random stars may, on its side, produce patterns that are not expected in real stars. 
With these further warnings in mind one can compare both histrograms for different values of $\ell$. As expected, $\ell=1$  results in too many cases with no polarization at all, something that has not been observed yet in Betelgeuse.  
On the other extreme, $\ell_{max}=6$ shows that two and three peaks are equally probable, contrary to observations that favour two peaks.

Our dataset contains two profiles with four peaks,  but we cannot conclude whether these are real or due to noise of some other bias.
This advices against considering $\ell_{max}$ larger than five.  
The right value of $\ell_{max}$ appears therefore  to be between $\ell_{max}=3$ and $\ell_{max}=5$.

\begin{figure}[htbp]
\includegraphics[width=.5\textwidth]{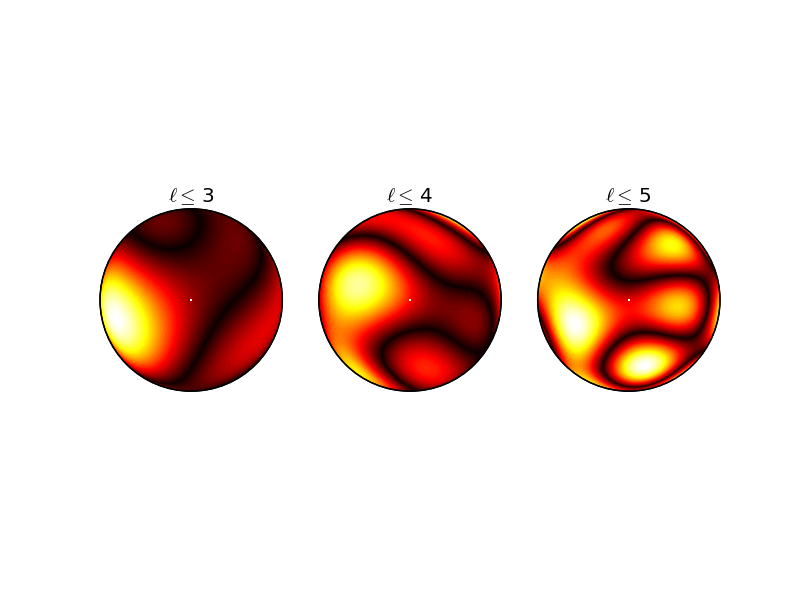}
\caption{Inversion of Betelgeuse profiles from January 9th, 2014, also shown in Fig. 1. 
From left to right the image has been made using only the spherical harmonic coefficients up to $\ell=3,4,$ and 5 respectively.  
As expected, the main spot appears from $l=3$ while higher coefficients keep adding more and more detail. }
\label{Y5_345}
\end{figure}

In order to decide between $\ell_{max}=3, 4,$ or 5 we run and show here two more tests. 
Figure\,\ref{Y5_345} shows the inversion for a given date with $\ell_{max}$=5. 
But only the right-most image shows the full result. 
In the leftmost image we show the reconstruction with just those coefficients up to $\ell=3$, and the centre image up to $\ell=4$. 
The main granule, here shown in the ambiguous solution that places it to the east, is the only structure recovered up to $\ell=3$. 
A transition image is seen in $\ell=4$ that adds small-scale structures which take  granular shape with $\ell=5$.
The series of images show that $\ell=3$ recovers the brightest granule, but that  $\ell=5$ is required to recover the right shape and size of all but the largest granule. The value $\ell=4$, while recovering also the smaller structures, corrupts their shape and produces elongated granules.

\begin{figure}[htbp]
\includegraphics[width=.5\textwidth]{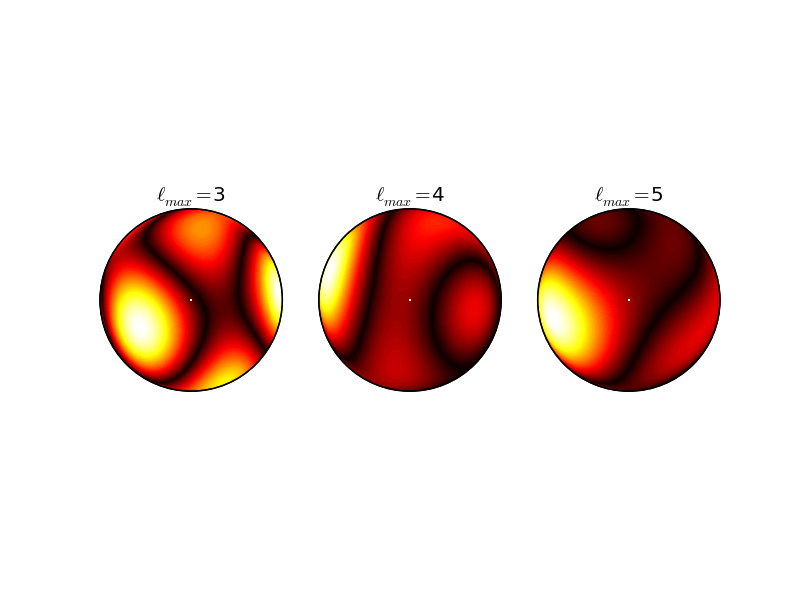}
\caption{Another different reconstruction of Betelgeuse on January 9th, 2014. 
This time the inversion with $\ell_{max}=3,4,$ and 5 are shown from left to right, but the reconstructions have been made using only the spherical 
harmonics coefficients up to $\ell=3$. 
This means that the left figure with $\ell_{max}=3$ is final for that inversion, but in the centre and right images the high order coefficients are 
missing from the reconstruction. 
The images $\ell_{max}=4$ and 5 show only the biggest granule, the fine details being missed with the high order coefficients. 
The $\ell_{max}=3$ case however tries to reproduce the profiles by adding large granules on the edges thus making them small by projection. 
This shows the necessity for $\ell_{max}>3$ to correctly invert the observed profiles.}
\label{Y345_3}
\end{figure}

A similar conclusion can be reached in the second and complementary test shown in Fig.\,\ref{Y345_3}. 
Here, three different inversions of the same data used in Fig.\,\ref{Y5_345} are presented with $\ell_{max}=3$ (left) to 5(right). 
But in each case, only the reconstruction up to $\ell=3$ is shown. 
This means that the image at left shows the final solution with $\ell_{max}=3$, but the centre and right images show only the result of the 
low-$\ell$ coefficients (spherical harmonics with $\ell=4$ and $\ell=4,5$ respectively are dropped). 
The main granule is seen in all three images, and we have used the 180-degrees ambiguity to place it always towards the east.
At $\ell=3$ the inversion tries to fit the rest of the information available in the Stokes profiles by adding granules at the limb. 
This can be seen as a strategy to make small structures with low $\ell$ : Although the full structure is large, placed at the limb, we only see one half or one quarter of it. 
This behaviour disappears as $\ell$ grows, the higher $\ell$ coefficients allow for the smaller structures and there is no need for the coefficients 
up to $\ell=3$ to reproduce the available information with granules at the limb, the coefficients of larger $\ell$  recover those small structures.

All these arguments combined together lead to the conclusion that $\ell_{max}=5$ recovers the actual size and number of the convective cells in Betelgeuse.
The typical size of convective cells for $\ell_{max}=5$ can be readily computed as $\pi/\ell=0.62$ times the disk radius, and their number in about 3 to 5 over the visible hemisphere of Betelgeuse

\section{Photospheric dynamics of  Betelgeuse during 4.5 years}

\begin{figure*}[htbp]
\includegraphics[width=\textwidth]{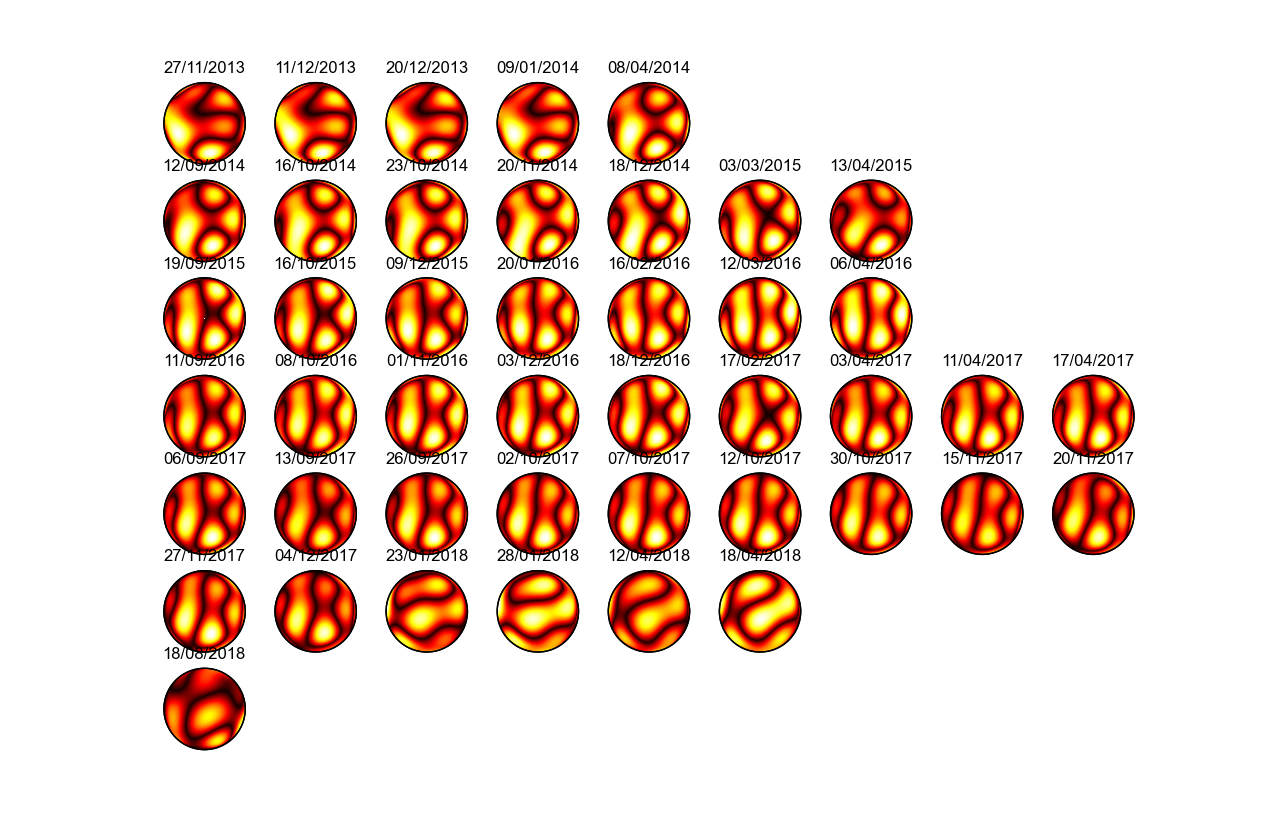}
\caption{Inferred images of Betelgeuse for the full dataset from November 2013 through August 2018, using  our new model with $\ell_{max}=5$. Brilliances are normalized on each image; the colour map is the same as that used in other figures, for example, in Fig. 3.}
\label{all}
\end{figure*}

Figure\,\ref{all} shows the inferred images of the convective cells in Betelgeuse from November 2013 to August 2018 in 44 observations.
The brightness is normalized for each image individually and all inversions have been made using $\ell_{max}=5$.  
The inversion process also takes advantage of the continuity of the observed profiles: the solution found for the previous date was used as initial 
solution of the next date in the Marquardt-Levemberg algorithm. 
Consequently, the 180 degrees ambiguity is solved once, in the first image and then the choice is carried on by continuity to all the other images.

\begin{figure*}[htbp]
\includegraphics[width=\textwidth]{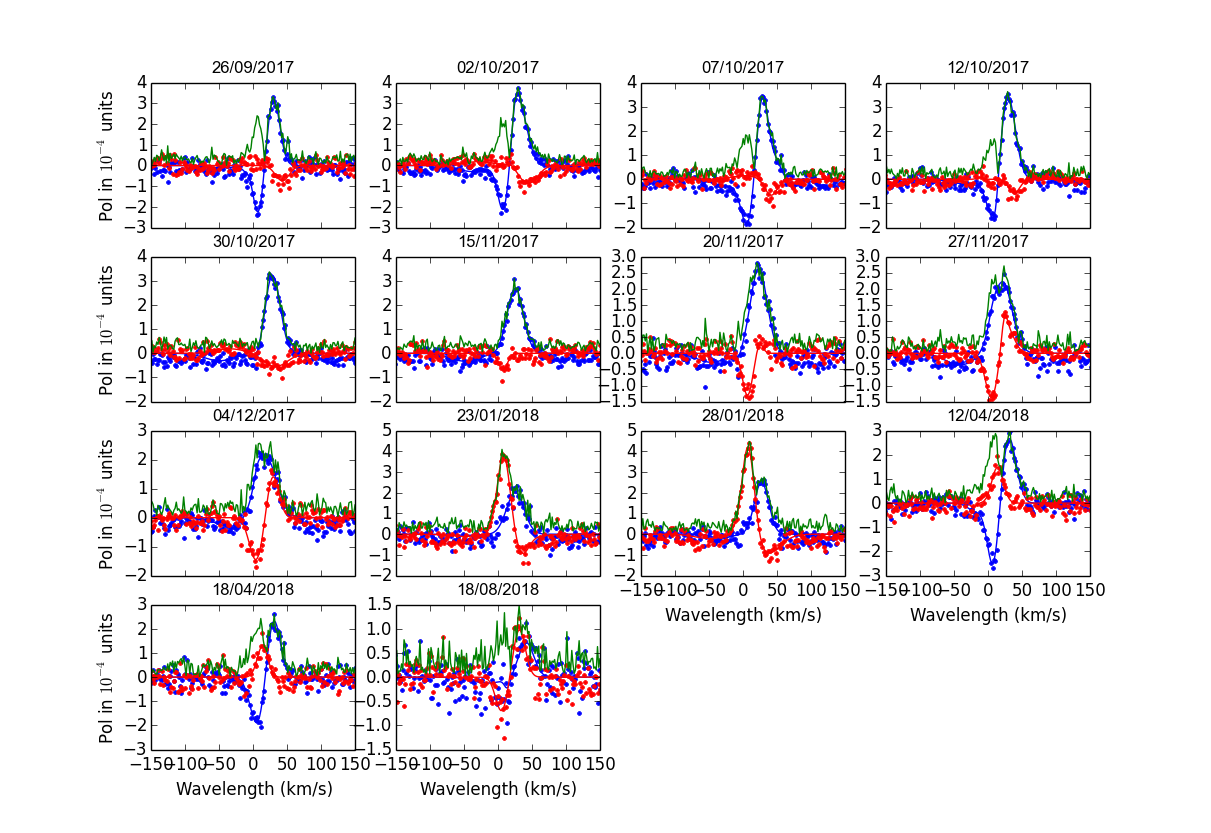}
\caption{Stokes $Q$ (blue) and $U$ (red) profiles observed at each night from September 2017 through August 2018. The total linear polarization profile $\sqrt{Q^2+U^2}$ is shown in green. We show only those data not present in \cite{Auriere2016} and \cite{Mathias2018}.}
\label{profiles}
\end{figure*}

Most of the data inverted for these images have already been presented by \cite{Auriere2016} (November 2013 -- April 2015) and by 
\cite{Mathias2018} (September 2015 -- April 2017). 
The data covering September 2017 -- August 2018 are used here for the first time. 
It  extends the initial monitoring already
presented in \cite{Auriere2016} and \cite{Mathias2018}. Fig. \ref{profiles} shows the actual LSD profiles for Stokes $Q$ and $U$ as well as the total lineal polarization $\sqrt{Q^2+U^2}$ for every observation date.

As in previous cases, observations were obtained with Narval on the Telescope Bernard Lyot at the Observatoire du Pic du Midi \cite{Auriere2003}. 
While the linear polarization signal is too small to be reliably detected in individual lines we have used the LSD technique \cite{Donati1997} to add-up the signal of several thousand lines (see details in \cite{Auriere2016}). 
The noise level on the resulting  LSD profile is a few times $10^{-5}$ the continuum intensity, while the peak amplitudes of linear polarization are at 
least ten times larger and reach, in some particular occasions, 0.1\,\% of the continuum intensity.

After inversion of all the available data set, we obtain a series of images dating back to the end of 2013.
The series starts with a dominating granule towards the eastern hemisphere and roughly centred on the equator. 
A secondary granule is seen in the western hemisphere but close to the South. 
As said before, the main granule coincides with the bright spot seen in interferometric data with PIONIER by \cite{Montarges2016} on January 2014 and also put 
in evidence by \cite{Auriere2016}.  
It dominates the inferred images of Betelgeuse until the end of 2014. 
During this period, it keeps moving around, stretching and coming back. 
There is a nice agreement between these motions and deformations and what numerical simulations show \cite[e.g.][]{Freytag2002,Chiavassa2011}.

By the end of 2014, the south  granule grows in importance and two other granules appear in the north-western quarter. 
This new scenario lasts until the end of 2016. 
At this point, we return to a situation similar to that of 2014, with the two granules, one in the southeast, and the other towards the south.  
This coming back suggests two possibilities.  
It could be that these convective cells are long-lasting structures of the atmosphere of Betelgeuse. 
Even when the star goes through a period of more vigorous convection, they keep their presence so that the star always shows at least the two of them. 
Another possibility is that during the period 2014-2016 these granules lost predominance but new granules in the north-west and north respectively took 
the leading role.
In such scenario, our choice between the two ambiguous solutions would be responsible for the granules being kept at the southern hemisphere.  
New interferometric observations could help us distinguish between such ambiguous solutions.

The main granule lasts for at least two years. If we take the images at face value, we see a small granule in the north-western part already in November 2013. 
It keeps growing in brightness till 2016 then dwindles back to the end of the series. 
This sets a lifetime of four years for the convective cells of Betelgeuse.  
\cite{Mathias2018} produced a temporal Fourier analysis of the signals and found considerable power for timescales around 2000 days. 
We find that these periods, that well correspond to the time scales associated to convection \cite{Stothers2010} are well represented by the evolution 
seen in our images. 
Nevertheless, it is also important to stress that the profiles and the images keep changing all the time. 
At intervals of one week to one month  the variations are well beyond noise levels. 
Although the number and average position of the granules are quite long-lasting, their actual shape and position keep changing in time spans of 
one month or even less, changes confirmed by the variability of the observed Stokes $Q$ and $U$ profiles.

We also searched for patterns in the time series of the harmonic coefficients. 
We do not see any indications of periods in individual coefficients. 
Some of them appear to show trends: they are either diminishing or growing during the full series. 
But in general we could not extract any clear information from individual coefficients.


\section{Conclusion}
We observe convective cells in the photosphere of Betelgeuse by inference from the inversion of linear polarization spectral profiles observed with the 
Narval instrument. 
The key to the interpretation is the blue-shift of the linear polarization signals, that we associate  with the brighter parts, those places where hot plasma rises. Simultaneously, the circular 
polarization signals, having its origin in the Zeeman effect, are attributed to weak magnetic fields. In analogy to the solar atmosphere, 
such weak fields are  dragged by convective flows to the dark intergranular 
lanes where the cold plasma sinks into the stellar interior. They will in consequence be redshifted with respect to the linear polarization signals.
This key fact is confirmed by a series of other results, as the timescales of the signals, the aspect of the bright structures seen and their dynamics.

We have been able to give a characteristic size of 0.6 stellar radii to these convective cells and to associate velocities to those plasma movements.
The characteristic upflow velocities measured are of the order of 20\,km.s$^{-1}$ and we also have a suggestion as to the values of the horizontal advection velocities that 
would be slightly smaller, but of the same order. 
Downflows velocities from  circular polarization are also of the order of 20\,km.s$^{-1}$ once the observed values are corrected from  projection onto the line of sight.

We have monitored Betelgeuse in linear polarization over more than four years. 
The shape and position of the convection cells change from month to month  and even from week to week, but individual structures can be tracked over years. 

Among the several flaws of the imaging method we developed and used throughout this work, one of the most obvious is the presence of ambiguities. We could rotate by 180 degrees every single image presented here and the linear polarization would fit identically well the observations. 
Because of this it is unclear at given times whether the chosen solution to this ambiguity problem is the right one. 
It is clear that periodic checks with other imaging methods, in particular interferometry, is desirable to fix these ambiguities.

\begin{acknowledgements}
This work was supported by the "Programme National de Physique Stellaire" (PNPS) of CNRS/INSU co-funded by CEA and CNES.
\end{acknowledgements}


\end{document}